\documentclass[aps, prb,
amsmath,amssymb,showpacs,groupedaddress]{revtex4}
\usepackage{graphicx}
\graphicspath{{../}}
\usepackage{epstopdf}
\begin{document}

\title{On the Simple Approach to Critical Phenomena Description.}

\author{A. S.~Yurkov}
\affiliation{644076, Omsk, Russia, e-mail: fitec@mail.ru}
\date{\today}


\begin{abstract}
{Simple field-theoretical approach to critical phenomena  is described. In contrast to the Wilson's theory, a description in real 3-dimensions space is used. At the same time the described approach is not the same as  Parisi's. Used subtraction scheme is different from the one used  by Parisi, but the main point is that we treat       changes of $T-T_C$ as explicit  perturbations. By such an approach not only the critical domain   but also  the crossover from critical domain  to domain of Landau theory  can be described  very simply.  Consideration is restricted to scalar $\phi^4$ model    in one-loop approximation.}
\end{abstract}

\maketitle


\section{Introduction}

The problem of critical phenomena description is known for a long time and was successfully solved in 70's mainly by Wilson \cite{bib:WilsonNL}. Wilson's theory is field-theoretical one and essentially use formal expansion in  series in $\varepsilon$,  so called $\varepsilon$-expansion. Here $\varepsilon$ is deviation of space dimensions from a case of 4 dimensions.

Besides Wilson's approach  using $\varepsilon$-expansion yet another field-theoretical  approach working directly in the real  3-dimensional space is also known. This approach  proposed by Parisi \cite{bib:Parisi73}  has been used in calculations of critical exponents of the scalar model up to five-loop approximation\cite{bib:BakerPRL,bib:BakerPR}. Similar approach was proposed by Ginzburg\cite{bib:Ginzburg}.

Up to now both these approaches have been developed in detail and now they form the standard basis of the critical phenomena theory  (see\cite{bib:Pelissetto2002,bib:ZJ-98,bib:ZJ00} and references therein).

However yet another field-theoretical approach to critical phenomena is possible. This approach is described in this paper.  As in  Parisi's approach we use renormalization group (RG)  in real 3-dimensional space. But in contrast to latter we use point of renormalization at arbitrary momentum. Such an arbitrary  point lets us to describe a massless field as well.  Besides, we treat  the  changes of $T-T_C$ as explicit perturbations  by additional term in  the Lagrangian.  Although such  details are not critical for physical meaning of a theory, these changes allow to simplify  essentially the physical sense of a theory. Besides that, such an approach allows to describe  very simply  not only  the critical domain  but also  the crossover from critical domain  to the  domain of Landau  theory.   It is worth to point out that the  aim of this paper is not to calculate critical exponents with high accuracy but to draw simple and intuitively clear physical picture of a critical phenomena. This is why the consideration is restricted to scalar $\phi^4$ model    in one-loop approximation and consideration is as simple  as possible.


\section{Perturbation theory and diagrams}

In this section we describe very briefly perturbation theory to introduce the notations. The details can be found in any quantum fields textbook, say by Ramond\cite{bib:Ramond}. 

Statistical physics of thermally fluctuating  scalar field $\phi$ with interactions described by term  $\phi^4$   is defined by partitional function which can be represented as path integral
\begin{equation}
\label{Z}
Z(J) = \int e^{-S(\phi , J)}{\cal D\phi} \, ,
\end{equation}
where "action" is
\begin{equation}
S(\phi , J)= S_0(\phi , J)+S_I(\phi)=\int {\cal L}_0 d^3{\bf x} + \int {\cal L}_I d^3{\bf x}\, ,
\end{equation}
and "Lagrangian" is
\begin{equation}
\label{L}
{\cal L}_0 =\frac{1}{2}(\nabla\phi)^2 + \frac{\tau_0}{2}\phi^2  -  J\phi \, , \qquad 
  {\cal L}_I =\frac{g_0}{4!}\phi^4  \,  .
\end{equation}
Generally speaking it should be additional $T^{-1}$ in the Gibbs exponent in (\ref{Z}) and some coefficient $f$ in $(\nabla\phi)^2 /2$ in (\ref{L}). But one can redefine field $\phi \to T^{-1/2}f^{1/2}\phi$ so that  these coefficients disappear.  Given such field redefinition $\tau_0\,$, $g_0$ and $J$ are also  redefined:
\begin{equation}
\tau_0 \to \tau_0 f^{-1} \, ,
\end{equation}
\begin{equation}
\label{g0}
g_0 \to g_0 T f^{-2} \, ,
\end{equation}
\begin{equation}
J \to J T^{-1/2} f^{-1/2} \, .
\end{equation}

It is easy to see that variational derivatives of $Z(J)$ by source field $J({\bf x})$ generates non-normalized correlators $G=Z(0)\langle \phi({\bf x}_1)\dots \phi({\bf x}_n)\rangle$ which can be named Green's functions in analogy with quantum field theory. To make path integral calculable, the exponent with the interaction term is expanded in series:
\begin{equation}
e^{-S_I} = \sum_n \frac{1}{n!}(-S_I)^n \, .
\end{equation} 
Such a representation yields the  well known perturbation theory which can be represented by Feynman diagrams. These diagrams consist of solid lines and elementary vertexes.  A solid line corresponds to the propagator (two-point Green's function) with  Fourier transform  $(k^2+\tau_0)^{-1}\,$. Elementary vertex is $-g_0/ (4!)\,$. 

Further one can introduce connected Green's functions $G_C$ and so called one participle irreducible Green's functions $\Gamma$ in a common way. Self-energy function (mass operator) $\Sigma(k)$ in a diagrammatic form is presented as a sum of two-point diagram (without external tails) which can not be divided in two parts by removing of one line. With self-energy function one can easily get the exact perturbated propagator: 
\begin{equation}
G_C({\bf k})=\frac{1}{k^2+\tau_0 - \Sigma(k)} \, .
\end{equation}
Other details see in the textbook cited above.


\section{Renormalization of interaction parameter}

Let us  describe 4-point one-participle irreducible Green's function $\Gamma_4\,$. It is  defined by diagram:

\begin{center}
\includegraphics[width=8cm,keepaspectratio]{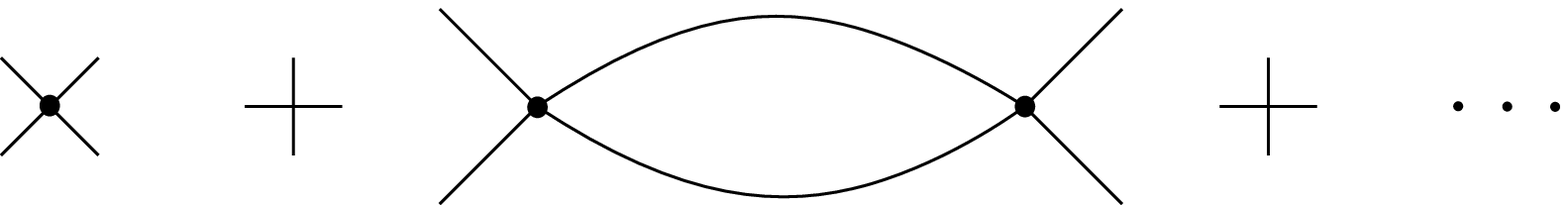}
\end{center}

\noindent
Analytical representation of this diagram is:
\begin{equation}
\label{G4-0}
-\Gamma_4({\bf p}_1, {\bf p}_2,  {\bf p}_3, {\bf p}_4) = g_0  - g_0^2 \sum I({\bf q})  \, ,
\end{equation}
where $I({\bf q})$ are loop integrals different from each other by permutations of diagram's tails;  ${\bf q}$ is the momentum which flows throw a loop,  it can be found from tail momenta ${\bf p}_a$  by momentum conservation law.

After that we renormalize the perturbation theory. Usually renormalization is used when loop integrals diverge. In three dimensions $\phi^4$ is super-renormalizable theory, there is the only divergent diagram and it is different from the shown above. But anyway it is possible to renormalize the convergent diagrams also.  Such a renormalization improves the perturbation theory. 

To renormalize perturbation theory we divide parameters into couples of terms $\tau_0=\tau+\tau_{ct}\,$, $g_0=g+g_{ct}\,$ and treat the term of Lagrangian with $\tau_{ct}$ as perturbation. Besides that,  $\phi^4$-terms with $g$ and $g_{ct}$ are treated as separate perturbations. Because the new perturbation series is series in $g\,$,  counterterms $\tau_{ct}\sim g$ and $g_{ct} \sim g^2$  at least. Generally speaking, the gradient term $(\nabla\phi)^2 /2$ should be divided into a main term and a counterterm as well. But  it is not necessary in one-loop approximation of $\phi^4$-theory and we do not describe it.

According to the above, renormalized equation (\ref{G4-0}) up to $g^2$ is:
\begin{equation}
\label{G4R}
-\Gamma_4({\bf p}_1, {\bf p}_2,  {\bf p}_3, {\bf p}_4) = g  - g^2 \sum I({\bf q})  + g_{ct} \, .
\end{equation}
Certainly there is a continuum of  variants to divide bare interaction constant $g_0$ into renormalized interaction parameter $g$ and counterterm $g_{ct}\,$. Thus we should add some renormalization conditions to choose one of them. 

Without perturbation theory it does not matter what variant to choose. But if one uses perturbations then some variants are much better then the other. Obviously perturbation expansion works well when third term in r.h.s. of (\ref{G4R}) approximately compensates the  second term. Thus it is convenient to choose renormalization condition as a condition of full compensation of the second and the third term when the  tail momenta are equal to the some four vectors, so called renormalization point vectors. Then the perturbation expansion will work well when the tail momenta are close to renormalization point vectors.

\begin{figure}[h]
\begin{center}
\includegraphics[width=4cm,keepaspectratio]{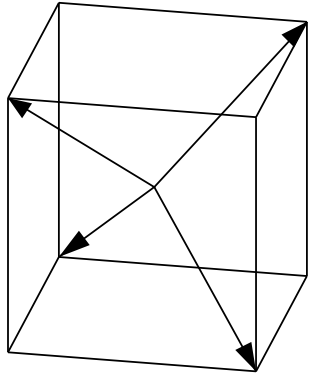}
\end{center}
%
\caption{Symmetric set of vectors ${\bf n}_a\,$. }
\label{fig:symnvectors}
\end{figure}

Now we should choose the normalization point. To parametrize four vectors by one parameter we introduce for unity vectors ${\bf n}_a\,$ and  define normalization point  as ${\bf p}_a=\mu {\bf n}_a\,$ where $\mu$ is renormalization parameter. It is convenient to choose  vectors ${\bf n}_a$  being inscribed in cube symmetrically as shown on FIG.\ref{fig:symnvectors}. Obviously with such vectors ${\bf n}_a$ vectors ${\bf p}_a=p{\bf n}_a$ obey momentum conservation low.

Further we restrict the consideration of $\Gamma_4({\bf p}_1, {\bf p}_2,  {\bf p}_3, {\bf p}_4 )$ to a case when ${\bf p}_a=p{\bf n}_a\,$.  Such a $\Gamma_4$ will be denoted as $\Gamma_4(p)\,$. For such a case ${\bf q}$ is the same for all permutations of tails and we can write: 
\begin{equation}
\label{Gp-ct}
-\Gamma_4( p) = g -
\frac{3}{2}g^2\frac{1}{(2\pi)^3}\int\frac{d^3{\bf k}}{(k^2+\tau)[({\bf k}+{\bf q})^2 +\tau ]} + g_{ct} \, .
\end{equation}
With elementary geometry one can find $q=(2p)/\sqrt{3}\,$. According to said above we define renormalized $g$ by condition:
\begin{equation}
\label{gdef}
\left .  -\Gamma_4(p)\right |_{p=\mu} = g  \, .
\end{equation}
Loop integrals similar to integral in (\ref{Gp-ct}) are usually  evaluated by a well known Feynman trick. Specifically this integral also can be calculated by elementary methods in cylindrical coordinates with z-axis along vector ${\bf q}\,$. Anyway integration and fixing of $g_{ct}$ by (\ref{gdef}) yields:
\begin{equation}
\label{Gp}
-\Gamma_4(p)=g -   g^2\frac{3\sqrt{3}}{16 \pi}
\left [ \frac{1}{p} \arctan\left(\frac{p}{\sqrt{3\tau}}\right) -
\frac{1}{\mu} \arctan \left(\frac{\mu}{\sqrt{3\tau}}\right) \right ]
\, .
\end{equation}

Actually equations (\ref{gdef}) and (\ref{Gp}) define how renormalized interaction parameter $g$ changes when  the scale of renormalization point $\mu\,$  changes. One need only substitute the new scale $\mu'$ as $p$ in (\ref{Gp}).  But it is worth to remember that (\ref{Gp}) works well only when $p$ is close to $\mu\,$. Thus it is much better to describe large change of $\mu\,$ as  step-by-step small changes of $\mu\,$. If each step is infinitesimal then this yields a differential equation.  This is the key idea of renormalization group method. Substituting $p=\mu+d\mu\,$ in (\ref{Gp}) with simple algebra we get:
\begin{equation}
\label{dgdmtau}
\frac{d g_{\mu}}{d \mu}= 
 g^2_{\mu}\frac{3\sqrt{3}}{16 \pi}\cdot\frac{1}{\mu^2}
\left[\arctan\left(\frac{\mu}{\sqrt{3\tau}}\right) -
\frac{\mu / \sqrt{3\tau} }{1 + ( \mu / \sqrt{3\tau} )^2}
\right] \, ,
\end{equation}
where the subscript $\mu$ stresses that $g$ is a function of $\mu\,$. This equation is one of the key equations of our description. When $\tau=0\,$ it becomes more simple:
\begin{equation}
\label{dgdm0}
\frac{d g_{\mu}}{d \mu}=  g^2_{\mu}\frac{3\sqrt{3}}{32}\cdot\frac{1}{\mu^2} \, .
\end{equation}
%


\section{Renormalization of "mass"}

The parameter $\tau$ corresponds to the square of the participle mass in quantum field theory. This parameter also undergoes renormalization.  In one-loop approximation this renormalization represented by single diagram for self-energy function:
\begin{center}
\includegraphics[width=1.5cm,keepaspectratio]{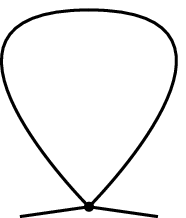}
\end{center}
It is essential that in one-loop approximation $\Sigma\,$ is momentum independent.  In our theory $\tau\,$ is $\mu$-independent constant while one-loop approximation is used.

It is worth to note that in the common theories $\tau$ is $\mu$-dependent nevertheless. This is reached by very refined trick which we do not use. Instead, we use another more physically clear trick which is described below.  


\section{"Running" interaction parameter}

In this section we solve equation (\ref{dgdmtau}) but  before that it is useful to describe solution of (\ref{dgdm0}). Obviously,  when $\mu>>\sqrt{\tau}\,$ the solution of (\ref{dgdmtau}) is close to the solution of (\ref{dgdm0}), while the latter is much more simple. Solution of (\ref{dgdm0}) is:
\begin{equation}
g = \frac{\mu}{A + C\mu} \, ,
\end{equation}
where $A=3\sqrt{3}/32\approx 0.1624\,$, and $C$ is an  integration constant. One can see easily that if $\mu \to \infty\,$ then $g \to C^{-1}\,$. Thus at large momentum  $g$ is constant. Let us denote this constant as $g_{\infty}\,$. Then
\begin{equation}
\label{gmut0}
g(\mu)=\frac{g_{\infty}}{1+Ag_{\infty}\mu^{-1}} \, .
\end{equation}
Otherwise if $\mu \to 0\,$ then
\begin{equation}
\label{gmto0}
g(\mu\to 0) =\frac{\mu}{A}\approx 6.158 \mu \, .
\end{equation}%

Equation (\ref{gmto0}) shows a very important fact: the behaviour of $g$ in IR limit is universal, it does not depend on $g_{\infty}\,$.  This means that if $\phi\,$ describes some field in real system, say in crystal, then interactions of its long-wave thermal fluctuations  does not depend on interactions at atomic scale. Now we see this fact for  $\tau=0\,$ (i.e. for a system at just critical point), further we will see this for $g_{\infty}^2>>\tau\ne 0\,$  as well.

Exact  solution of (\ref{dgdmtau}) yields not too complex result:
\begin{equation}
g=\frac{\mu}{[(3\sqrt{3})/(16\pi)]\arctan(\mu/\sqrt{3\tau}) + C\mu} \, ,
\end{equation}
where $C$ is an  integration constant. We see again that $g \to {\rm const}=C^{-1}=g_{\infty}\,$ while
$\mu \to \infty\,$.  So we can rewrite the equation in the form:
\begin{equation}
\label{gmgen}
g(\mu)=\frac{g_{\infty}}{1 + g_{\infty}\mu^{-1}[(3\sqrt{3})/(16\pi)]\arctan(\mu/\sqrt{3\tau})} \, .
\end{equation}

Some essential point should be discussed  here. Describing solution of differential equation mathematically,  one may think that the  integration constant $C\,$ and the  parameter $g_{\infty}\,$ can depend on $\tau\,$. But by physical reasons  this parameters should be  $\tau$-indepedent. Indeed, one can see from (\ref{Gp}) that if $\mu\to \infty,$ then $g_{ct}\to 0\,$. Thus $g_{\infty}\,$ actually is bar parameter $g_0\,$ which obviously  does not depends on $\tau\,$.

Further,  if $\mu = 0\,$ then:
\begin{equation}
g(\mu=0)=\frac{g_{\infty}}{1 + g_{\infty}[3/(16\pi)]\tau^{-1/2}} \, .
\end{equation}
One can see very easily that if $\tau<<g^2_{\infty}\,$ then the behaviour is universal again:
\begin{equation}
g(\mu=0) = \frac{16\pi}{3}\sqrt{\tau} \approx 16.755 \sqrt{\tau}\, .
\end{equation}
Besides if $g_{\infty}>>\mu >> \sqrt{\tau} \,$ then
\begin{equation}
g(\mu)= \frac{32}{3\sqrt{3}}\mu  \approx 6.158 \mu  \, .
\end{equation}
This equation is also universal and it is the same that was derived above for the massless ($\tau=0$) field.


\section{Perturbation of  proximity to critical point}

Experimentally, the proximity of a system to the  critical point is usually  controlled by temperature. Thus a direct way to describe system reaction to the changes of proximity to critical point requires to find a reaction to $g_0$ changes (see equation (\ref{g0})~). But if $T-T_C<<T_C\,$ then it does not matter what to change, $T$ or $T_C\,$. This is why the proximity to critical point can be controlled by $\tau_0\,$ changes, this is more convenient technically.

We describe $T-T_C$ changes as an explicit perturbation. Thus we add a perturbation term $t \lambda_0 \phi^2/2 \,$ to the "Lagrangian".  Here the parameter $t$ controls the changes of $T-T_C\,$. Why a parameter $\lambda_0$ is added will be clear further. 

In reality, $t$ is spatially homogeneous. However, perturbation with  spatially homogeneous $t$ can not be renormalized directly. This is why we generalize the theory to spatially inhomogeneous $t({\bf x})\,$ which will be changed to homogeneous $t$ at the end of calculations. Actually we add yet another source field interacting  with $\phi\,$ non-linearly.

Perturbation $t({\bf x}) \lambda_0 \phi^2/2 \,$ generates additional 3-tail vertex ($-\lambda_0/2\,$ in the analytical representation). One of its tails is $t({\bf x})\,$ which will be depicted  in the diagrams as a waving line.  As any vertex,  $\lambda$-vertex can be renormalized. To renormalize it we divide it into renormalized vertex $\lambda$ and counterterm $\lambda_{ct}$ where $\lambda_{ct} \sim g$ at least. Further descriptions are generally the same as the renormalization of $g\,$.  One-loop correction to $\Gamma_3$ is represented by diagram:
\begin{center}
\includegraphics[width=5cm,keepaspectratio]{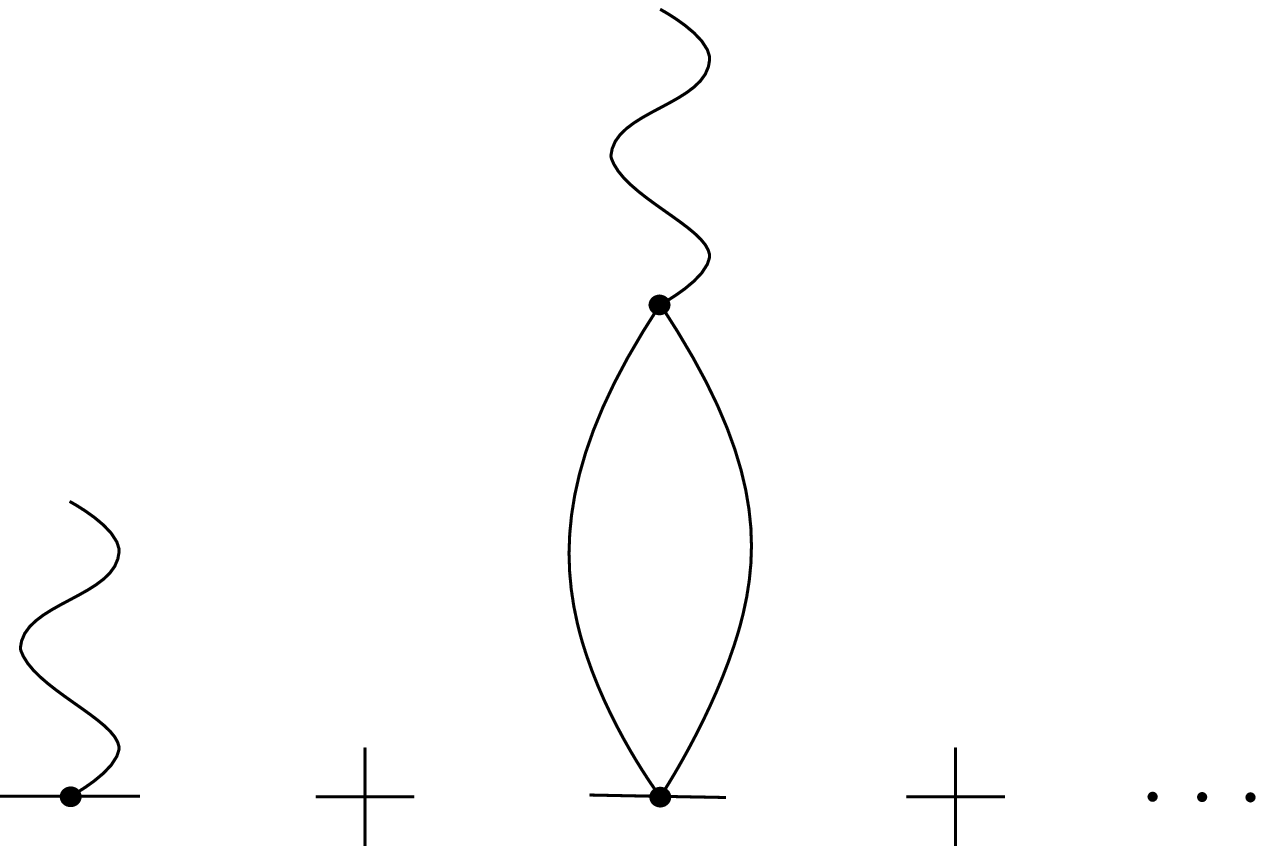}
\end{center}
which analytically yields:
\begin{equation}
\label{G3I}
-\Gamma_3(p)=\lambda  - \frac{1}{2}g\lambda \frac{1}{(2\pi)^3}\int
\frac{d^3{\bf  k}}{(k^2+\tau)[({\bf  k}+{\bf  p})^2+\tau]} + \lambda_{ct} \, ,
\end{equation}
where ${\bf p}\,$ is waving tail momentum. Renormalization conditions we choose similar to condition for $g\,$:
\begin{equation}
\left . -\Gamma_3(p) \right |_{p=\mu} = \lambda \, .
\end{equation}
Integral in (\ref{G3I}) is just the same that in (\ref{Gp-ct}). So with full analogy to above descriptions we derive:
\begin{equation}
-\Gamma_3(p) = \lambda - \lambda g \frac{1}{8\pi} \left[
\frac{1}{p}\arctan\left(p / \sqrt{4\tau}\right) - \frac{1}{\mu}\arctan\left(\mu / \sqrt{4\tau}\right) 
\right] \, , 
\end{equation}
\begin{equation}
\label{dlgen}
\frac{d \lambda_{\mu}}{d \mu} = \lambda_{\mu} \frac{g_{\mu}}{8\pi}\cdot\frac{1}{\mu^2}\left[
\arctan\left(\mu / \sqrt{4\tau}\right) - \frac{\mu / \sqrt{4\tau}}{1 + \left(\mu / \sqrt{4\tau}\right)^2}
\right] \, .
\end{equation}
For $\tau=0\,$ or $\mu>>\sqrt{\tau}\,$ the last equation can be simplified:
\begin{equation}
\label{dlt0}
\frac{d \lambda_{\mu}}{d \mu} = \lambda_{\mu} \frac{g_{\mu}}{16}\cdot\frac{1}{\mu^2} \, .
\end{equation}

\section{"Running" $\lambda$}

$\lambda(\mu)\,$ can be found easily  when $\tau=0\,$ or $\mu>>\sqrt{\tau}\,$. Using (\ref{gmut0}) and(\ref{dlt0}) we obtain:
\begin{equation}
\lambda(\mu) = C \left(\frac{\mu}{\mu + g_{\infty}\left(3\sqrt{3}/32\right)}\right)^{\alpha} \, ,
\end{equation}
where $\alpha=2/(3\sqrt{3})\approx 0.385\,$, $C\,$  is an  integration constant. Obviously if $\mu\to\infty\,$  then $\lambda \to C={\rm const}\,$. Thus the equation can be rewritten as:
\begin{equation}
\label{lmt0}
\lambda(\mu) = \lambda_{\infty}
\left(\frac{\mu}{\mu + g_{\infty}\left(3\sqrt{3}/32\right)}\right)^{\alpha} \, .
\end{equation}
Here the parameter $\lambda_{\infty}\,$ is $\tau\,$-independent by the same reason as $g_{\infty}\,$.
When $\mu<< g_{\infty}\,$ the equation can be simplified:
\begin{equation}
\label{lmt0asim}
\lambda(\mu) = \frac{\lambda_{\infty} \mu^{\alpha}}
{\left[ g_{\infty}\left(3\sqrt{3}/32\right)\right]^{\alpha}}
 \, .
\end{equation}

If $\mu \sim \sqrt{\tau\,}$ or $\mu <\sqrt{\tau}$ then the differential equation (\ref{dlgen})  apparently can be solved by numerical methods only.  But at least in critical domain it is easy to get universal correlation between  $\lambda(\mu=0)$ and $\lambda(\mu_*)\,$, where $\mu_* \sim \sqrt{\tau}\,$, say $\mu_*=10\sqrt{\tau}\,$. Further $\lambda(\mu_*)\,$ can be evaluated by (\ref{lmt0}) or (\ref{lmt0asim}). 

If $\mu<\mu_*=10\sqrt{\tau}\,$ and $\mu_*<<g_{\infty}\,$ then (\ref{gmgen}) can by simplified as
\begin{equation}
\label{gmgensimp}
g(\mu)=\frac{16\pi\mu}{3\sqrt{3}\arctan(\mu/\sqrt{3\tau})} \, .
\end{equation}
With such $g(\mu)\,$ equation (\ref{dlgen}) yields:
\begin{equation}
\label{llaster}
\int\limits^{\lambda(\mu_*)}_{\lambda(0)} \frac{d\lambda}{\lambda} =
\alpha \int\limits_0^{10}\frac{\left[\arctan(x/2) - (x/2)/[1+(x/2)^2] \right]} 
{x\arctan(x/\sqrt{3})} dx \, ,
\end{equation}
where substitution $x=\mu/\sqrt{\tau}\,$ used.

In (\ref{llaster}) r.h.s. is a universal constant which can be evaluated numerically. For our purpose however  it does not matter what   its value is. The only fact we need is that according to (\ref{llaster})
\begin{equation}
\lambda(0) \sim \lambda(\mu_*) \, .
\end{equation}
With (\ref{lmt0asim}) this correlation yields in critical domain:
\begin{equation}
\label{l0rdy}
\lambda(0) \sim \tau^{\alpha/2} \, .
\end{equation}


\section{Critical exponents}

The last equation of the previous section allows us to derive critical exponents very simply. If the system is  off the critical point by some non-zero $\tau\,$,  then, in order  to bring  it to  the critical point, the perturbation of the  self-energy function $\Sigma(k=0)\,$ should be equal $\tau\,$. On the other hand,  in the first order of perturbation theory $\Sigma=-\lambda(0)\Delta T\,$, where   $\Delta T\,$ is the temperature shift (negative  to shift down to critical temperature). Thus we have a self-consistent equation:
\begin{equation}
\tau=-\lambda(0)\Delta T \, .
\end{equation}
Using (\ref{l0rdy}) we instantly obtain:
\begin{equation}
\tau \sim \tau^{\alpha/2} (T-T_C) \, .
\end{equation}
An obvious  transformation yields:
\begin{equation}
\tau \sim (T-T_C)^{\gamma} \, ,
\end{equation}
where
\begin{equation}
\gamma = \frac{2}{2-\alpha} \, .
\end{equation}
There is no field renormalization in one-loop approximation. This is why  the  Fisher exponent $\eta\,$ is zero and the  propagator is $G(k)=(k^2+\tau)^{-1}\,$. Thus the  critical exponent for correlation length is
\begin{equation}
\nu=\frac{\gamma}{2} = \frac{1}{2-\alpha} \, .
\end{equation} 
Numerical estimations yields values  $\gamma=1.24\,$, $\nu=0.62\,$ which are close to the  known precise values   (see\cite{bib:Pelissetto2002}) .  

Derivations in this section above  may look doubtful because $\lambda(0)\,$ changes while $\tau\,$ changes. Using of the first order perturbation may also seem doubtful. But we can justify the result by consideration of infinitesimal change of $\tau\,$ with infinitesimal change of temperature. Such an approach yields:
\begin{equation}
d \tau = \lambda(0) d T \sim \tau^{\alpha/2} d T \, .
\end{equation} 
Integration of this differential equation yields exactly the same result that we  have obtained before.


\section{Crossover to Landau theory}

\begin{figure}[h]
\begin{center}
\includegraphics[width=12cm,keepaspectratio]{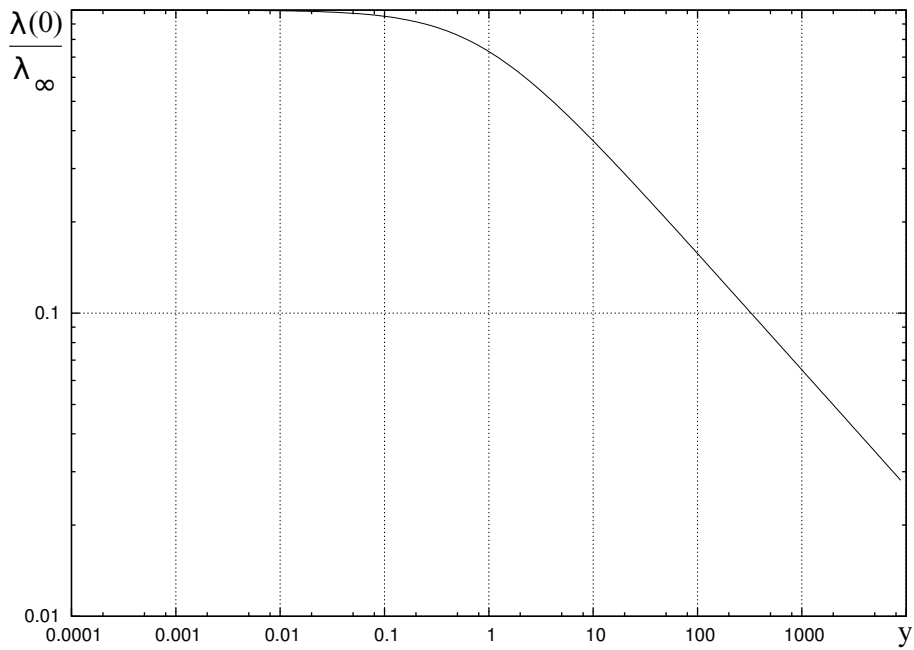}
\end{center}
%
\caption{Dependence of $\lambda(0)/\lambda_{\infty}\,$ on $y=g_{\infty}/(8\pi\sqrt{\tau})\,$ calculated numerically.  }
\label{fig:lont}
\end{figure}

The approach considered here allows us  to determine not only critical behaviour but also crossover to Landau theory behaviour when $\tau\,$ is large. To describe this crossover one should find $\lambda(0)\,$ as a function of $\tau\,$ without approximations.  Using (\ref{gmgen}) and (\ref{dlgen}) we obtain
\begin{equation}
\lambda(0) = \lambda_{\infty} e^{-F(y)} \, ,
\end{equation}
where function $F(y)\,$ is defined as
\begin{equation}
F(y) = y\int\limits_0^{\infty} 
\frac{\arctan(x/2)-(x/2)/[1+(x/2)^2] }
{x + y\alpha^{-1}\arctan(x/\sqrt{3})} 
\cdot \frac{dx}{x} \, ,
\end{equation}
$y=g_{\infty}/(8\pi\sqrt{\tau})\,$, $x=\mu/\sqrt{\tau}\,$.
A dependence of $\lambda(0)/\lambda_{\infty}\,$ on $y=g_{\infty}/(8\pi\sqrt{\tau})\,$  can be calculated by numerical methods. The result is shown on FIG.\ref{fig:lont}.

From FIG.\ref{fig:lont} one can see that if $\sqrt{\tau} << g_{\infty}/(8\pi)\,$ i.e $y >> 1\,$ then $\lambda(0)\,$ depends on $\tau\,$ as described in previous sections. This behaviour  corresponds to critical domain.  Otherwise if $\sqrt{\tau} >> g_{\infty}/(8\pi)\,$ i.e.  $y << 1\,$ then $\lambda(0)\approx \lambda_{\infty} = {\rm const}\,$. This means that in this domain $d\tau \sim dT\,$. Such a behaviour corresponds to  Landau theory. One can see also that condition of crossover $\tau \sim g_{\infty}/(8\pi)\,$ i.e. $y \sim 1\,$ coincides, at least up to some coefficient, with well known Levanyuk criterion.


\section{Conclusion}

In this paper a very simple approach to critical behaviour description has been  presented. Generally this approach is close to the  common field-theoretical theories using renormalization group method. Nevertheless the  presented approach has at least one advantage: it allows one to draw a simple, intuitively clear physical picture of a critical phenomena.  

Physical reason why non-trivial exponents  appear is that the  system's  sensitivity to $T-T_C\,$ variations, described by  $\lambda$-vertex at zero momentum, changes as system goes toward to critical point. If one takes in account this fact, then reasonable values of non-trivial exponents  can be obtained by very simple calculations. Moreover, far from critical point $\lambda$-vertex ceases  to  depend on correlation length (which is $\sim 1/\sqrt{\tau}\,$). In this way,  far from critical point the behaviour of a system becomes identical to the one described by Landau theory. Such a crossover can also be described in a simple way.


\end{document}